\documentstyle[preprint,aps]{revtex}

\ifpreprintsty
\preprint{IMSc-97/03/07}
\begin{document}
\draft 
\title{Solar neutrinos : Eclipse effect}
\author
{Mohan Narayan, G. Rajasekaran, Rahul Sinha} 
\address
{Institute of Mathematical Sciences, Madras 600 113, India.}
\date{7 March 1997}
\maketitle

\else

\oddsidemargin=-6pt
\textwidth=6.5in
\begin{document}
\begin{flushright}
IMSc-97/03/07
\end{flushright}

\vspace{5ex}
\begin{center}

{\LARGE \bf Solar neutrinos : Eclipse effect}\\
\bigskip
\bigskip
{\Large Mohan Narayan, G. Rajasekaran, Rahul Sinha}\\
\bigskip
{\it Institute of Mathematical Sciences, Madras 600 113, India.}\\
\bigskip
(\today)
\end{center}
\vspace{1in}
}
\fi
\begin{abstract}

It is pointed out that  the enhancement of the solar
neutrino rate in  a real time detector like Super--Kamioka, SNO or
Borexino  due to neutrino oscillations in the moon during a partial or total
solar eclipse may be observable. The
enhancement is calculated as a function of the neutrino parameters in
the case of  three flavor mixing. This  enhancement
if seen, can further help to determine  the neutrino parameters.
 
\end{abstract}
%\end{document}

\newpage
\section{Introduction}

The sun is a copious source of neutrinos with a wide spectrum of energies
and these neutrinos have been detected by terrestrial neutrino detectors,
although at a rate lower than expected from theoretical calculations.
A new generation of detectors \cite{totsuka,sno,raghavan} with high counting rates will
soon be producing abundant data on solar neutrinos.
Mixing and the consequent oscillations among the neutrinos of different
flavors is generally believed to be the cause of the reduced intensity of
neutrino flux detected on the earth. However, neutrino--oscillation is a
complex phenomenon depending on many unknown parameters (six parameters for
three flavors  $\nu_{e}$, $\nu_{\mu}$, $\nu_{\tau}$) and considerable amount
of experimental work and ingenuity will be required before the neutrino
problem is solved.

Hence it would be desirable, if  apart from direct detection, we can
subject the 
solar neutrino beam to further tests by passing it through different
amounts of matter, in our attempts to learn more about the neutrinos. Nature
has fortunately provided us with such opportunities:
$(1)$ Neutrinos detected at night pass through the earth.
$(2)$ Neutrinos detected during a solar eclipse pass through the moon.
$(3)$ Neutrinos detected at the far side of the earth during a solar
eclipse pass through the moon and the earth. We shall call this
scenario $(3)$ a
{\em double eclipse}. The scenario $(1)$ has been studied in the literature
rather extensively \cite{blwz,ours}. The purpose of the present work is to examine
the scenarios $(2)$, and $(3)$. Two previous works \cite{currsc,Sher} have
discussed scenario $(2)$, however both are incomplete in many
respects.

The plan of the paper is as follows. The relevant astronomy
is presented in Sec.\ref{ecl}. In Sec.\ref{th} we give the theory of the passage of the
solar neutrinos through the moon and the earth, taking into account properly
the non-adiabatic transitions occuring at the boundaries of the moon
and the earth. In Sec.\ref{calc} we present the
numerical calculations of the neutrino detection rates during the single and
double eclipses and the results. Sec.\ref{disc} is devoted to discussion.

\section{Eclipses and double eclipses}
\label{ecl}

Solar neutrinos are produced within the solar core whose radius is of
order $1/10$ of the solar radius and we shall approximate this by a point
at the centre of the sun. What is required for our purpose is that the lunar
disc must cover this point at the centre of the sun and so as far as the
neutrino radiation is concerned , the solar eclipse is more like an 
occultation of a star or a planet by the moon.

Astronomers characterize the solar eclipse by the optical coverage $C$
which is defined as the ratio of the area of the solar disc covered by 
the lunar disc to the total area of the solar disc.For neutrino physics we
require the distance $d_M$ travelled by the neutrino inside the moon.
Defining the fraction $x =\displaystyle\frac {d_M} {2 R_M}$ where $R_M$ is the lunar radius, $x$
can  be given in terms of $C$ by the following formulae:
\begin{eqnarray}
x  &=& \sqrt {(4 z (2 - z) - 3)}\label{eq1} \\
C &=& \frac {2} {\pi} \left(\cos^{-1} (1- z) - (1- z) \sqrt{z (2 - z
)}\right) 
\label{eq2}
\end{eqnarray}
Eqn(\ref{eq2}) can be inverted to get the parameter $z$ for a given
$C$ and this $z$ can be substituted in eqn(\ref{eq1}). The
relationship between $x$ and $C$ so obtained is plotted in Fig\ref{Fig. 1}. When
the lunar disc passes through the centre of the sun, $C$ is 0.39 and
the neutrino eclipse starts at this value of $C$.When the optical
coverage increases above 39\%, $x$ rises sharply from zero and reaches
0.6 and 0.95 for optical coverage of 50\% and 80\% respectively.

	For any point of observation of the usual solar eclipse (which
we shall call {\em single eclipse})  there
is a corresponding point on the other side of the earth where a double
eclipse occurs. With the coordinates labeled as (latitude, longitude), 
the single eclipse point ($\alpha,\beta$) is related to the double
eclipse point ($\lambda,\sigma$) by the relations (see Fig.\ref{Fig. 2}):
\begin{eqnarray}
\lambda&=&\alpha+2 \delta\nonumber \\
\sigma&=&\pi-2\:\Theta_{UT}-\beta %\;(mod 2\pi),
\end{eqnarray}
where,
\begin{equation}
\sin \delta = \sin 23.5^o \sin (\frac{2\pi t}{T_{Y}}) ,
\end{equation}
$T_Y$ is the length of the year , zero of time t is chosen at midnight of
autumnal equinox {\it i.e.}Sept. 21, and $\Theta_{UT}$ is the angle
corresponding to the Universal Time -- UT.

During a double eclipse, the neutrinos  travel through the earth 
in addition. The distance $d_E$ travelled by the 
neutrino inside the earth along
the chord between the points $(\alpha,\beta)$ and
$(\lambda,\sigma$)  as a function of time $t$, is given by 
\begin{equation}
 d_E = 2 R_E (\sin \lambda \sin \delta + \cos \lambda \cos\delta \cos
(\frac{2\pi t}{T_D}))
\end{equation} 
where $R_E$ is the radius of the earth and $T_D$ is the length of the
day. This is the same distance that needs to be calculated in the
study of the day--night effect and a plot of this distance as a
function of $t$ is given in our earlier paper\cite{ours}.

 Present and upcoming high statistics neutrino detectors expect to
collect a few solar neutrino events every hour.  As discussed in
Secs.~\ref{moon} and \ref{secmoonearth}, single and double eclipse can
lead to enhancements of rates by upto two and a half times. Even with
such large enhancements during the eclipse the signal may not exceed
statistical errors, since each solar eclipse lasts only for a few
hours. However they occur fairly often. As many as 32
solar eclipses are listed to occur during the 14 year period 1996
through 2010.  Global maps and charts are available\cite{http} for
location and duration of both the umbral and penumbral coverage.
	
	The eclipses during the 2 year period 1997 through
1999 are the following:
\begin {description}
\item [1997 March 9] -- Total Solar Eclipse
\item [1997 Oct 12] -- Partial Solar Eclipse
\item [1998 Feb 26] -- Total Solar Eclipse
\item [1998 Aug 22] -- Annular Solar Eclipse
\item [1999 Feb 16] -- Annular Solar Eclipse
\item [1999 Aug 11] -- Total Solar Eclipse
\end{description}
We have analyzed the five total/annular eclipses. We have not examined
the partial solar eclipse of 1997 October 12, since we have not so far
been able to procure the data for this eclipse from the references
cited in \cite{http}. It is important to remark here that although
partial solar eclipses are not so useful to astronomers they are
nevertheless relevant for neutrino physics as long as $C$ is above
$0.39$.

A study of the global maps of the five total/annular eclipses listed
above shows that only three of them can occur as a single or a double
eclipse at any of the three existing detector sites Kamioka, Sudbury and Gran
Sasso with coordinates $(36.4^oN,140^oE),\;(46.5^oN,81^oW)$ and
$(42.5^oN,13.5^oE)$ respectively. Approximate estimates are presented
below  for the duration 
and optical coverage of the relevant detectors at these sites:
\begin{description}
\item [1997 March 9] -- This eclipse will be a single eclipse
for the Super--Kamioka, with an approximate duration of two and a half
hours and a maximum optical coverage of just over 60\%. In addition,
there will be a double eclipse at the Gran--Sasso, the site of Borexino,
at almost the same time as at Kamioka and for the same duration, with a
70\%  optical coverage.
\item [1998 Feb 26] -- Though Sudbury will have an optical eclipse the
coverage is less than 39\% and so no neutrino--eclipse occurs. However, a
double eclipse occurs at Gran--Sasso with a maximum coverage of 70\%
to 80\%. This corresponds to the single eclipse at ($31^oN,45^oW$).
\item [1999 Aug 11] -- This eclipse will provide one of the best
opportunities, as a single eclipse with 90\% optical coverage and almost 3 hour duration
at Gran-Sasso. In addition Super--Kamioka site will also get a double
eclipse with 90\% to 100\% optical coverage, corresponding to the
single eclipse at approximately ($36^oN,40^oE$)
\end{description}

\section{Theory}
\label{th} 
\subsection{Regeneration in the moon}
\label{moon}
We now describe a straightforward way of obtaining the neutrino
regeneration effect in the moon by using a model of moon of constant density
($3.33$gms/cc). 
Let a neutrino of flavor $\alpha$ be produced at time $t=t_0$ in the
core of the sun.
Its  state vector is 
\begin{equation}
| \Psi_{\alpha} (t_0) \rangle = |\nu_\alpha \rangle = \sum_i U^S_{\alpha i}
| \nu_i^S \rangle.
\end{equation}
where $|\nu_i^S \rangle$ are the matter dependent mass eigenstates
with mass eigenvalues $\mu_i^S$ and
$U^S_{\alpha i}$ are the matrix elements of the matter dependent mixing
matrix in the core of the sun. We use Greek index $\alpha$ to denote the three
flavors e, $\mu$,$\tau$ and Latin index i to denote the mass eigenstates
i = 1,2,3. The neutrino propagates in the sun adiabatically upto $t_{R}$ (the
resonance point), makes non-adiabatic transitions at $t_R$, propagates
adiabatically upto $t_1$ (the edge of the sun) and propagates as a free particle
upto $t_2$ when it enters the moon. So the state vector at $t_2$ is
\begin{equation}
| \Psi_{\alpha} (t_2) \rangle 
= 
\sum_{j,i}  |\nu_j \rangle exp \left(-i \varepsilon
_{j} (t_{2}- t_{1}) \right)  exp \left(-i\int_{t_{R}}^{t_1} \varepsilon^{S}_{j} (t) dt 
\right) M^S_{ji} exp \left(-i \int_{t_0}^{t_R} \varepsilon^S_{i}(t) dt \right) 
U^S_{\alpha i}.
\label{psi2}
\end{equation}
where $ \varepsilon^S_{i} (t)(\equiv E+(\mu_i^S(t))^2/{2 E})$ are the matter dependent energy eigenvalues in the sun,
$ \varepsilon_{i}$ and $|\nu_{i} \rangle$ are the energy eigenvalues and the 
corresponding eigenstates in vacuum and $M^S_{j i}$ is the probability amplitude for
the non-adiabatic transition $i\rightarrow j$. We multiply the right hand side of  
eq.(\ref{psi2}) by $ \sum_{k} |\nu^M_{k} \rangle \langle \nu^M_{k}|$ ( = 1)where 
$| \nu^M_{k}\rangle$ (i =1,2,3) is the complete set of matter dependent mass 
eigenstates inside the moon. The neutrino propagates upto the the other end of the
moon at $t_3$, and the  state vector at $t_3$ is
\begin{eqnarray}
| \Psi_{\alpha} (t_3) \rangle & = & \sum_{k,j,i} |\nu_{k}^M \rangle
exp 
\left( -i \varepsilon_{k}^M (t_{3}-t_{2})\right) 
\langle \nu_{k}^M|\nu_{j} \rangle 
exp 
\left( -i \varepsilon_{j} (t_{2}-t_{1}) -i \int_{t_R}^{t_1}
\varepsilon^S_{j} (t) dt \right)\nonumber\\  \nonumber
& &\times M^S_{j i} exp \left(-i \int_{t_0}^{t_R} \varepsilon^S_{i}
(t) dt \right) U^S_{\alpha i}\\  \nonumber 
&=& \sum_{k,j,i} |\nu_{k}^M \rangle
exp 
\left( -i \varepsilon_{k}^M (t_{3}-t_{2})\right) 
M^M_{k j} 
exp 
\left( -i \varepsilon_{j} (t_{2}-t_{1}) -i \int_{t_R}^{t_1}
\varepsilon^S_{j} (t) dt \right)\\  
& &\times M^S_{j i} exp \left(-i \int_{t_0}^{t_R} \varepsilon^S_{i}
(t) dt \right) U^S_{\alpha i}
\label{psi3}
\end{eqnarray}
We have introduced the probability amplitude $M^M_{k j}$ for non-adiabatic transitions
$j\rightarrow k$ due to the abrupt change in density when the neutrino enters the 
moon . It is given by
\begin{equation}
  M^M_{k j} = \langle \nu_{k}^M| \nu_{j} \rangle = \sum_{\gamma} \langle \nu_{k}^M|
\nu_{\gamma} \rangle \langle \nu_{\gamma}| \nu_{j}\rangle
    =   \sum_{\gamma} U^M_{\gamma k} U^*_{\gamma j}
\label{noref}
\end{equation}
where $U_{\gamma j}$ is the mixing matrix in vacuum .
We multiply the right hand side of eq.(\ref{psi3}) by  $ \sum_{l} |\nu_ {l} \rangle \langle \nu_{l}|$ 
where  $| \nu_ {l}\rangle$ (l =1,2,3) is the complete set of vacuum mass eigenstates.
The neutrino leaves the other end of the moon at $t = t_3$ and propagates upto the
surface of the earth ,which it reaches at $t_4$ .So the state vector at $t_4$ is
\begin{eqnarray}
| \Psi_{\alpha} (t_4) \rangle & = & \sum_{k,j,i,l} |\nu_{l} \rangle 
exp 
\left( -i \varepsilon_{k}^M (t_{3}-t_{2})\right) 
\langle \nu_{l}|\nu_{k}^{M} \rangle M^{M}_{k j} 
exp 
\left( -i \varepsilon_{j} (t_{2}-t_{1}) -i \int_{t_R}^{t_1}
\varepsilon^S_{j} (t) dt \right) \nonumber\\
& &\times M^S_{j i} exp \left(-i \int_{t_0}^{t_R} \varepsilon^S_{i}
(t) dt \right) U^S_{\alpha i} exp \left(-i \varepsilon_{l}
(t_{4}-t_{3})\right) \nonumber\\
%&=& \sum_{k,j,i,l |\nu_{l} \rangle 
%exp 
%\left( -i \varepsilon_{k}^M (t_{3}-t_{2})\right) 
%M^{M}_{k j} M^{M}_{l k} 
%exp 
%\left( -i \varepsilon_{j} (t_{2}-t_{1}) -i \int_{t_R}^{t_1}
%\varepsilon^S_{j} (t) dt \right)\\  \nonumber
% &\times M^S_{j i} exp \left(-i \int_{t_0}^{t_R} \varepsilon^S_{i}
%t) dt \right) U^S_{\alpha i} exp \left(-i \varepsilon_{l}
%(t_{4}-t_{3})\right)\\ \nonumber 
&=& \sum_{k,j,i,l} |\nu_{l} \rangle
M^{M}_{k j} M^{M *}_{k l} M^{S}_{j i}
U^{S}_{\alpha i} exp \left (-i \Phi_{i j k l} \right)
\label{psi4}
\end{eqnarray}
where
\begin{equation}
\Phi_{i j k l} = 
\varepsilon_{k}^M (t_{3}-t_{2}) +\varepsilon_{l}(t_{4}-t_{3}) +\varepsilon_{j}
(t_{2}-t_{1}) + \int_{t_R}^{t_1} \varepsilon^{S}_{j} (t) dt + \int_{t_0}^{t_R}
\varepsilon^S_{i} (t) dt  
\end{equation}
We have used the fact that the the probability amplitude  for non-adiabatic 
transitions $k \rightarrow l$
 due to the abrupt change in density when the neutrino leaves the 
moon   is 
\begin{equation}
 \langle \nu_{l}| \nu_{k}^{M} \rangle = M^{M*}_{k l}
\end{equation}
%where $U_{\rho l}$ is the mixing matrix in vacuum .
The probability of detecting a neutrino of flavor $\beta$ at $t_{4}$ is
\begin{equation}
|\langle \nu_\beta |\Psi_{\alpha} (t_4) \rangle|^2 = \\ \nonumber 
\sum \\ \nonumber
U^{*}_{\beta l} U_{\beta l'} M^M_{k j} M^{M*}_{k' j'} M^{M*}_{k l} M^{M}_{k' l'} M^S_{j i} M^{S*}_{
j' i'} U^S_{\alpha i} U^{S*}_{\alpha i'}  \\ \nonumber 
exp \left(-i(\Phi_{i j k l} -
\Phi_{i' j' k' l'} \right)
\end{equation}
where the summation is over  the set of indices $i,j,k,l,i',j',k',l'$ 
Averaging over $t_{R}$ leads to $\delta_{i i'} \delta_{j j'}$ and this
results in the desired incoherent mixture of mass eigenstates of
neutrinos reaching the surface of the moon at $t_2$. Calling this
averaged probability as $P_{\alpha \beta}^M$ ( the probability for a
neutrino produced in the sun as $\nu_\alpha$ to be detected as
$\nu_\beta$ in the earth after passing through the moon), we can write the result as
\begin{equation}
 P_{\alpha \beta}^M = \sum_{j} P^S(\alpha \rightarrow j) P^M(j
\rightarrow \beta) 
\label{PabM}
\end{equation}
where
\begin{eqnarray}
 P^{S}(\alpha \rightarrow j) & = & \sum_i |M^S_{j i}|^2 |U^S_{\alpha
i}|^2  \\
P^M(j \rightarrow \beta) & = & 
\sum_{l,k,l',k'} U^{*}_{\beta l} U_{\beta l'} M^M_{k j} 
M^{M*}_{k' j} M^{M*}_{k l} M^{M}_{k' l'} exp \left(-i(\varepsilon^M_{k} - \varepsilon^M_{k'})
d_M - i(\varepsilon_{l} - \varepsilon_{l'}) r \right)
% & = &\sum_k \sum_{k'} \sum_{\gamma} \sum_{ \gamma'} U^{E*}_{\beta k} 
% U^E_{\beta k'} U^E_{\gamma k} U^{*}_{\gamma j} U^{E*}_{\gamma' k'}
% U_{\gamma' j} exp \left (-i(\varepsilon^E_{k}- \varepsilon^E_{k'}d
% \right)	
\end{eqnarray}
where we have replaced $(t_{3}- t_{2})$ by $d_{M}$, the distance travelled by the neutrino
inside the moon, and $(t_{4} -t_{3})$ by $r$  the distance travelled by the neutrino 
from the moon to the earth.
If there is no moon , we put $d_M$ = 0, so that 
$P^M(j \rightarrow \beta)$ becomes $|U_{\beta j}|^2$ and so eqn.(\ref{PabM})
reduces to the usual \cite{parke,narayan} averaged  probability 
for $\nu_{\alpha}$ produced in
the sun to be detected as $\nu_{\beta}$ in the earth :
\begin{equation}
P^{O}_{\alpha \beta} = \sum_{i,j} |U_{\beta j}|^2 |M^S_{j i}|^2
|U^{S}_{\alpha i}|^2 .
\label{PabO}
\end{equation}

\subsection{Regeneration during double eclipse.}
\label{secmoonearth}

We use a model of earth of constant density (5.52gms/cc). We start
with $\Psi_\alpha(t_4)$ given by eq.(\ref{psi4}) and multiply the right hand side by $\sum_p |\nu_{p}^E \rangle \langle \nu_{p}^E |$
 (= 1) where $|\nu_{p}^E \rangle$ $(i = 1,2,3)$  is a complete set of mass eigenstates 
inside the earth. The neutrino enters the earth at time $t$ = $t_{4}$ and is detected
at time $t$ = $t_{5}$ inside the earth.The state vector at time $t$= $t_{5}$ is
\begin{equation}
| \Psi_{\alpha} (t_5) \rangle  =  \sum_{k,j,i,l,p} |\nu_{p} \rangle
M^{E}_{p l} M^{M}_{k j} M^{M *}_{k l} M^{S}_{j i}
U^{S}_{\alpha i}  \exp \left (-i \Phi_{i j k l p } \right)
\end{equation}
where
we have introduced the probability amplitude $M^E_{p l}$ for non adiabatic transitions
$l\rightarrow p$ due to the abrupt change in density when the neutrino enters the 
earth . It is given by
\begin{equation}
 M^E_{p l} = \langle \nu_{p}^E| \nu_{l} \rangle = \sum_{\sigma} U^E_{\sigma p} U^*_{\sigma l}
\end{equation}
%where $U_{\rho l}$ is the mixing matrix in vacuum .
and
\begin{equation}
\Phi_{i j k l p } = 
\varepsilon_{p}^{E} (t_{5}-t_{4})+ \varepsilon_{k}^M (t_{3}-t_{2}) +\varepsilon_{l}(t_{4}-t_{3}) +
\varepsilon_{j} (t_{2}-t_{1}) + \int_{t_R}^{t_1} \varepsilon^{S}_{j} (t) dt + \int_{t_0}^{t_R}
\varepsilon^S_{i} (t) dt  
\end{equation}
The probability of detecting a neutrino of flavor $\beta$ at $t_{5}$ is
\begin{eqnarray}
|\langle \nu_\beta |\Psi_{\alpha} (t_5) \rangle|^2 &=& 
\sum 
U^{E*}_{\beta p} U^{E}_{\beta p'} M^{E}_{p l} M^{E*}_{p' l'} 
M^M_{k j} M^{M*}_{k' j'} M^{M*}_{k l} M^{M}_{k' l'} M^S_{j i} M^{S*}_{
j' i'} U^S_{\alpha i} U^{S*}_{\alpha i'} \nonumber \\ 
&&\times\exp \left(-i(\Phi_{i j k l p} -
\Phi_{i' j' k' l' p'} \right)
\end{eqnarray}
where the summation is over  the set of indices $i,j,k,l,p,i',j',k',l'p'$ 
Again averaging over $t_{R}$  and
calling this
averaged probability as $P_{\alpha \beta}^{M E}$ ( the probability for a
neutrino produced in the sun as $\nu_\alpha$ to be detected as
$\nu_\beta$ in the earth after passing through the moon and the earth), we can write the result as
\begin{equation}
P_{\alpha \beta}^{M E} = \sum_{j} P^S(\alpha \rightarrow j) P^{M E} (j
\rightarrow \beta) 
\label{PabME}
\end{equation}
where
\begin{eqnarray}
%P^{S}(\alpha \rightarrow j) & = & \sum_i |M^S_{j i}|^2 |U^S_{\alpha
%i}|^2  \\
P^{M E}(j \rightarrow \beta) & = & 
\sum_{l,k,p,l',k',p'} U^{E*}_{\beta p} U^{E}_{\beta p'} 
M^E_{p l} M^{E*}_{p' l'}  
M^M_{k j} M^{M*}_{k' j} M^{M*}_{k l} M^{M}_{k' l'} \nonumber \\
&&\times exp \left(-i(\varepsilon^E_{p} - \varepsilon^E_{p'}) d_{E} 
-i(\varepsilon^M_{k}-\varepsilon^M_{k'}) d_{M}
- i(\varepsilon_{l} - \varepsilon_{l'}) r \right)
\label{moonearth}
\end{eqnarray}
where we have replaced $(t_{5}- t_{4})$ by $d_{E}$, the distance travelled by the neutrino
inside the earth, $(t_{3}- t_{2})$ by $d_{M}$, the distance travelled by the neutrino
inside the moon and $(t_{4} -t_{3})$ by $r$  the distance travelled by the neutrino 
from the  moon to  the earth.

For the sake of completeness, we state that if we put $d_{M}$ = 0 in
eqn(\ref{moonearth}) 
we get the regeneration in earth alone:
\begin{equation}
P_{\alpha \beta}^E = \sum_{j} P^S(\alpha \rightarrow j) P^E(j
\rightarrow \beta) 
\label{PabE}
\end{equation}
where
\begin{equation}
P^E(j \rightarrow \beta)  =  
\sum_{k,k'} U^{E*}_{\beta k} U^{E}_{\beta k'} M^E_{k j} 
M^{E*}_{k' j} exp \left(-i(\varepsilon^E_{k} - \varepsilon^E_{k'})
d_{E} \right)
\label{Pejb}
\end{equation}
Eqns(\ref{PabE}) and (\ref{Pejb}) have been used to study the day--night effect \cite{ours}.

It is important to note that the factorization of probabilities seen
in eqs(\ref{PabM}),(\ref{PabME}) and (\ref{PabE}) is valid only for
mass eigenstates in the intermediate state. An equivalent statement of
this result is that the density matrix is diagonal only in the
mass-eigenstate representation and not in the flavour representation.

\subsection{Three flavor mixing parameters}
We parameterize the mixing matrix U in vacuum as
$U = U^{23} (\psi) U^{13} (\phi) U^{12} (\omega)$
where $U^{ij} (\theta_{ij})$ is the two flavor mixing matrix between the $i^{th}$ and the
$j^{th}$ mass eigenstates with the mixing angle $\theta_{ij}$, neglecting CP violation.
In the solar neutrino problem $\psi$ drops out \cite{ajmm,kuopan}  
The mass differences in
vacuum are defined as $\delta_{21} = \mu^{2}_{2} - \mu_{1}^{2}$ and
$\delta_{31} = \mu^2_{3} - \mu^2_{1}$. It has been shown \cite{narayan,fogli} 
that the
simultaneous solution of both the solar and the atmospheric neutrino
problems requires
\begin{equation}
\delta_{31} \gg \delta_{21}
\label{delgt}
\end{equation}
 and under this condition $\delta_{31}$ also drops out.
The rediagonalization of the mass matrix in the presence of matter (in
solar core, moon or
earth) under condition (\ref{delgt}) leads to the following results \cite{narayan}
\begin{eqnarray}
\tan 2 \omega_m  &=&  \frac{\delta_{21} \sin 2 \omega}{
		  \delta_{21} \cos 2 \omega - A \cos^2 \phi} \label{eq13} \\ 
\sin \phi_{m} &=& \sin \phi  \label{eq14} \\
\delta_{21}^{m} &=& \delta_{21} \cos 2 (\omega - \omega_{m})- A \cos^2\phi \cos 2 
\omega_{m} 
\label{eq15}
\end{eqnarray} 
where A is the Wolfenstein term $A = 2 \sqrt{2}~G_F~N_e\,E$ ($N_{e}$
is the number density of electrons and E is the neutrino energy)
%in and A is in $MeV^{2)})$
. We note that $\delta_{31}\gg A^S, A^M, A^E$. In
eqs.(\ref{eq13})--(\ref{eq15}) the subscript ``$m$'' stands for
matter. Under the condition $\delta_{31}\gg A\approx\delta_{21}$ we
need the non adiabatic transition probability $|M^{S}_{ij}|^2$ for i,j
=1,2 only and this is taken to be the modified Landau--Zener jump
probability for an exponentially varying solar density
\cite{kuopan}.

\section{calculations and Results}
\label{calc}
The neutrino detection rates for a Super--Kamoika type of  detector is
given by
\begin{equation}
R =\int\phi(E)\,\sigma (E) P_{ee} dE +
\frac{1}{6} (\int\left( \phi (E) \sigma (E) (1-P_{ee}) dE \right) 
\end{equation}
where the second term 
is the  neutral current contribution and $\phi(E)$ is the solar
neutrino flux as a function of the neutrino energy $E$ and
$\sigma(E)$ is the cross section from neutrino electron scattering and
we integrate from  $5MeV$ onwards. The
cross section is taken from  \cite{bahcall} and the flux from
\cite{bahpin92}. The rates for a single eclipse, double eclipse and
without eclipse (at day--time) $R_{M},R_{ME}$ and $R_{O}$  are calculated
using $P_{ee}^M,P_{ee}^{ME}$ and $P_{ee}^O$ from
eqns.(\ref{PabM}),(\ref{PabME}) and (\ref{PabO}) respectively. 
We define the enhancement factors $F$ and $G$ for a single and double
eclipse respectively:
\begin{eqnarray}
%\frac{N-D}{N+D}\equiv \frac{(R_{N}- R_{D})}{(R_{N} + R_{D})}.
F &=& \frac{R_{M} - R_{O}} {R_{O}}\\
G &=& \frac{R_{ME} - R_{O}} {R_{O}}.
\end{eqnarray}

It is easy to see that $F$ and $G$ have to be less than 5 and this
theoretical maximum value occurs when $P_{ee}^O=0$ and $P_{ee}^M$ and
$P_{ee}^{ME}$ are put 1. If one imposes the constraint that the
observed \cite{mohanetal} neutrino rate is $0.51\pm 0.07$ of the
prediction of the 
standard solar model, the maximum possible  enhancement is
reduced to about 1.40 (at 90\% C.L.).

We calculate the enhancement
factors $F$ and $G$  for various values of the neutrino
parameters, $\omega$, $\delta_{21}$, and $\phi$.
We show the results as  contour plots in the $\delta_{21}$--$\omega$ plane for
different values of  $\phi$. Figs.\ref{Fig. 3} and \ref{Fig. 4} show
the  $F$--contours for $\phi=0^o$ and $\phi=30^o$ respectively. 
For each $\phi$ we show the contours for different
distances of travel of the neutrino through the moon.
Fig. \ref{Fig. 5} shows the $G$--contours for $\phi=0$ for the maximum
distance of travel of the neutrino inside the moon and the earth.
The main features of the results are as follows:

\begin{itemize}
\item As the distance travelled by the neutrino inside the moon increases one can
see an appreciable increase in the enhancement factor $F$. It increases from less than 
$10\%$ to about almost $100\%$
when the neutrino travels the whole diameter of
the moon in the case of two flavor mixing i.e $\phi$ = $0$.
\item Large  ($> 40\%$) values of $F$ occur for
$\omega$ between $20^o$ and $30^o$ and $\delta_{21} \sim 10^{-6} eV^2$
This is true even for nonzero $\phi$.
\item The effect of a non zero ``13'' mixing angle $\phi$, is to dilute the enhancement
factor $F$ for all values of distance travelled through the moon.(In fact for $\phi \approx
45^o$, $F$ is practically zero and so we have not plotted this case.)
This is because
a non zero $\phi$ means $\nu_{e} \leftrightarrow \nu_{\tau}$ oscillations,
and matter cannot  reconvert $\nu_{\tau}$ back to $\nu_{e}$, because the
``13'' mixing angle $\phi$ is not affected by matter.
\item If large enhancement $F$ is seen for values of $x \leq 0.6$, it immediately
signals a very small value of $\phi$. On the other hand, if no enhancement is seen for
small $x$ but  there is enhancement
only for $x \geq 0.8$ it signals an appreciable value of $\phi$.
\item For a double eclipse there are considerable enhancements even for small
values of $\omega$.  There is enhancement throughout the range of $\omega$
from small angles till about $40^o$. In fact the regions of largest enhancement ($>100\%$) 
are for $\omega$ between $5^o$ to $20^o$.
\item The region of maximum enhancement factor $G$ is centered around a value of
$\delta_{21}$ which is a little above  $10^{-6} eV^2$, this being the value for maximum in $F$. 
This can be traced to the
fact : $A^{E} > A^{M}$. However sizeable enhancement occurs over a wide range of
$\delta_{21}$.
\item If enhancement is not seen, then certain regions of the neutrino parameter space can
be excluded.If no enhancement is seen for single eclipse, a panel of $\omega$ between 
$5-25^o$ and
$\delta_{21} \approx 2 \times 10^{-7}-2 \times 10^{-6} eV^2$ for $\phi$ = 0 can be ruled
out. If it is not seen for a double eclipse, a larger region can be ruled out.

\end{itemize}

\section{Discussion}
\label{disc}
We have studied the effect on the solar neutrinos of their passage through
the moon as well as the moon together with the earth. Although the
numerical results presented in the paper cover only a representative 
sample of the set of various parameters, our analytical expressions can be
used for more extensive calculations depending on the requirement. Also
one can go beyond the hierarchy : $\delta_{31} >> \delta_{21}$.

We now offer a few concluding remarks:
\begin{itemize}
\item Together with the day-night effect, the eclipse effects provide us
with the tools for studying solar neutrinos, in a way independent of
the uncertainties of the solar models.
\item If the neutrino mass differences are really very small
($\delta_{21} < 10^{-5} eV^2$) there is no way of pinning down the neutrino
parameters  other than using the astronomical objects such as the moon or
the earth for the "long-base-line experiments".
\item It is important to stress that even the demonstrated absence of any
eclipse effect would provide us with definitive information on neutrino
physics.
\item Accumulation of data over many eclipses may be needed for good statistics.
In any future planning of detector sites, this may be kept in mind.
\item It appears that Nature has chosen the neutrino parameters in such a
way that the sun affects the propagation of solar neutrinos. It may be hoped
that Nature has similarly chosen "lucky" parameters so that the moon and the
earth too can affect  the neutrinos!
\item Finally, we stress the novelty of the whole phenomenon, and urge the
experimentalists to look for and study the eclipse effects in an unbiased 
manner. They may even discover some surprises, not predicted by our
calculations!
\end{itemize}

Acknowledgments: We thank KVL Sarma for bringing ref \cite{Sher} to our attention,
N.D. Hari Dass for raising the possibility of detecting the neutrinos on the opposite
side of the earth during the eclipse, and M.C Sinha for much help with the astronomical
aspects of the problem and Sandip Pakvasa for useful discussions.

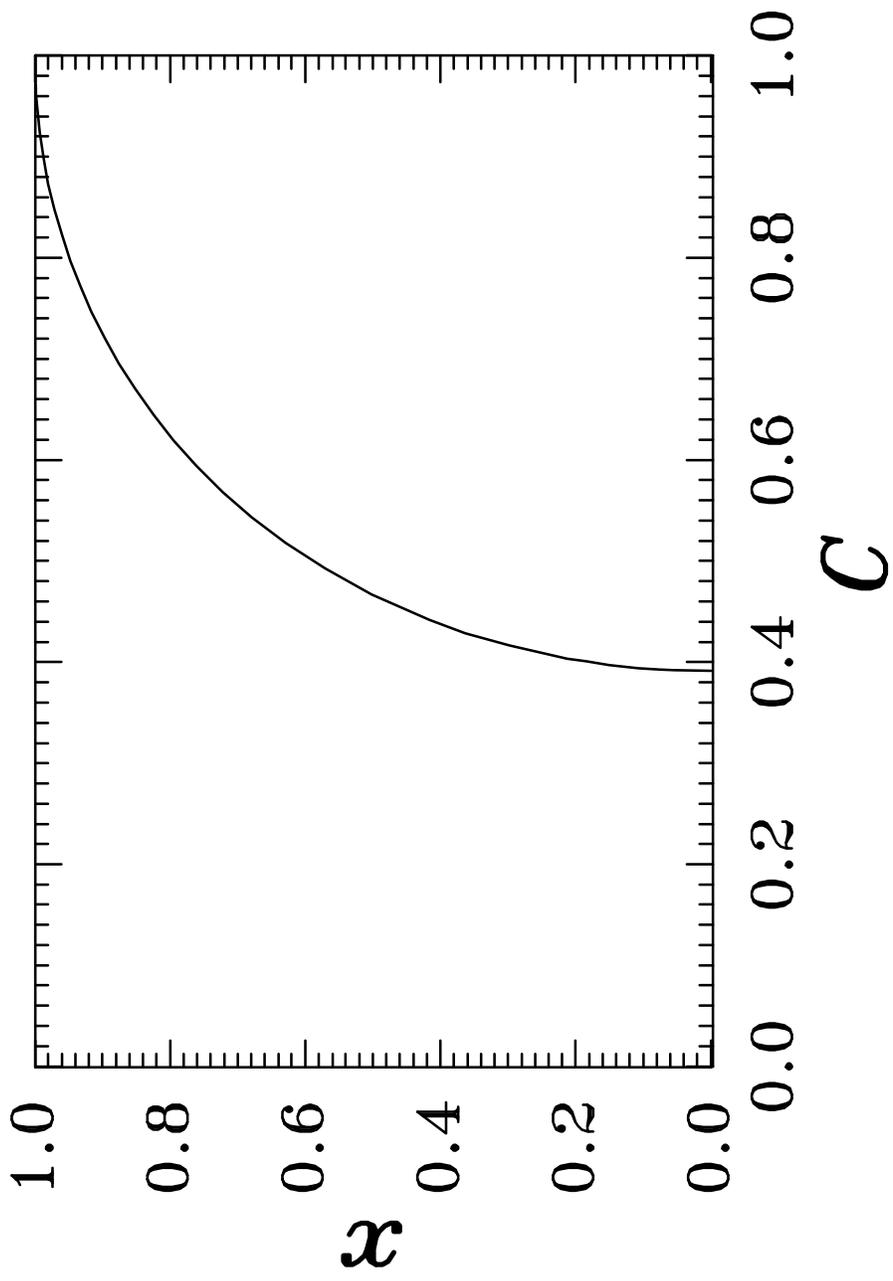
\begin{figure}
\caption{
 The fractional distance travelled by the neutrino inside the moon
$x (= \displaystyle \frac {d_{M}} {2 R_{M}}$) is plotted against the optical
coverage $C$ of the solar eclipse.
}\label {Fig. 1}
\end{figure}
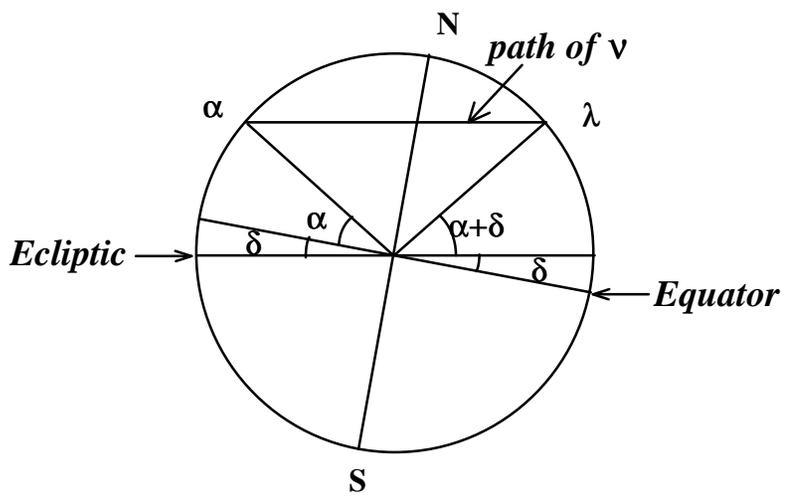
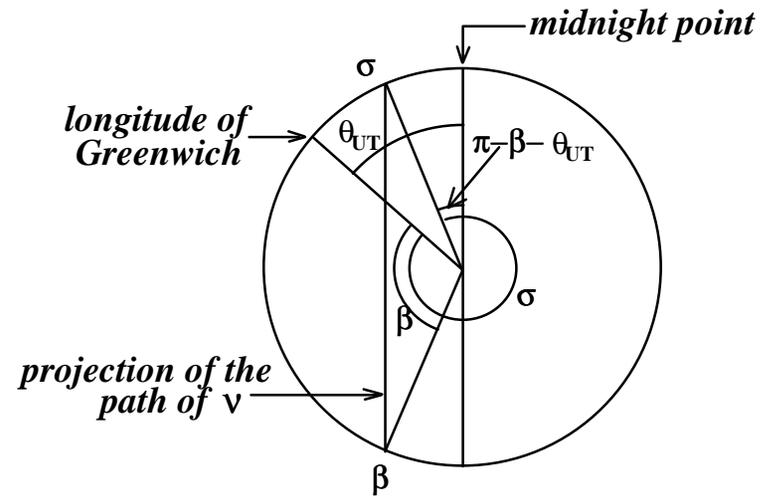
\begin{figure}
\caption{Geometry relating the double eclipse point $(\lambda,\sigma)$
to the single eclipse point $(\alpha,\beta)$. $(a)$ Section of the earth passing
through $(\alpha,\beta)$ and perpendicular to the ecliptic. $(b)$
Section passing through $(\alpha,\beta)$ and parallel to the equator.
}\label {Fig. 2}
\end{figure}
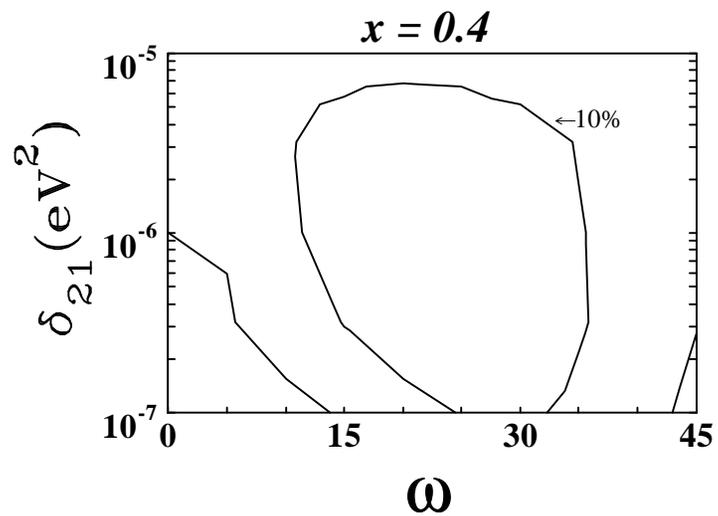
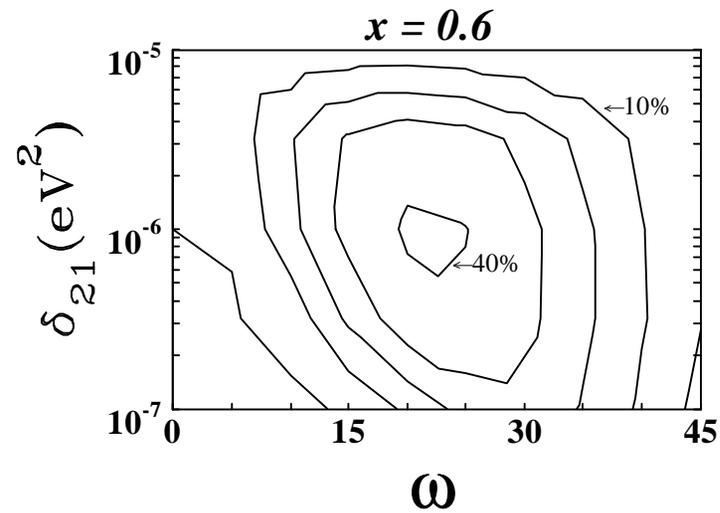
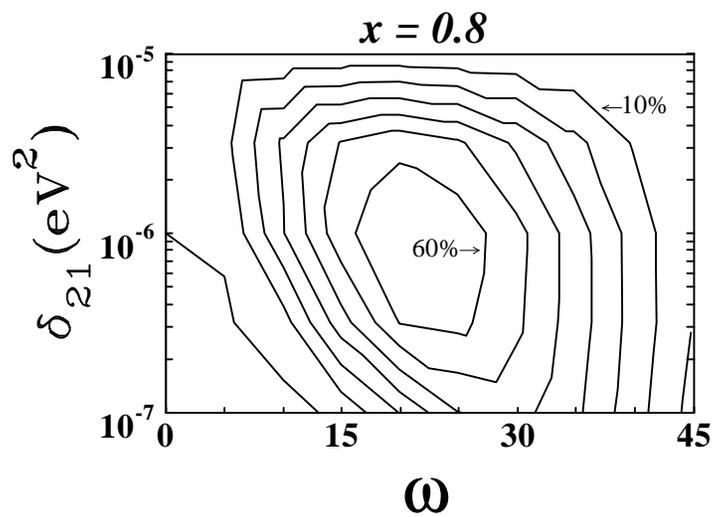
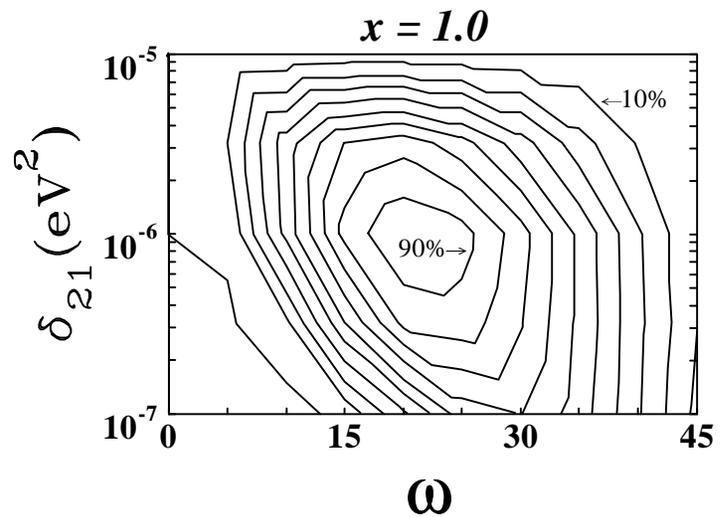
\begin{figure}
\caption{
 Contour plots of the enhancement factor for single eclipse
$F( = \displaystyle\frac {R_{M} - R_{O}} {R_{O}})$ in the $\omega - \delta_{21}$ plane
for $\phi = 0^o$ and for four values of $x$ ($x$ = 0.4, 0.6, 0.8 and 1.0).
The enhancement factor (regarded as a percentage) increases by $10\%$ for
every adjacent ring , as we move inwards towards the centre of the plot.
}\label {Fig. 3}
\end{figure}
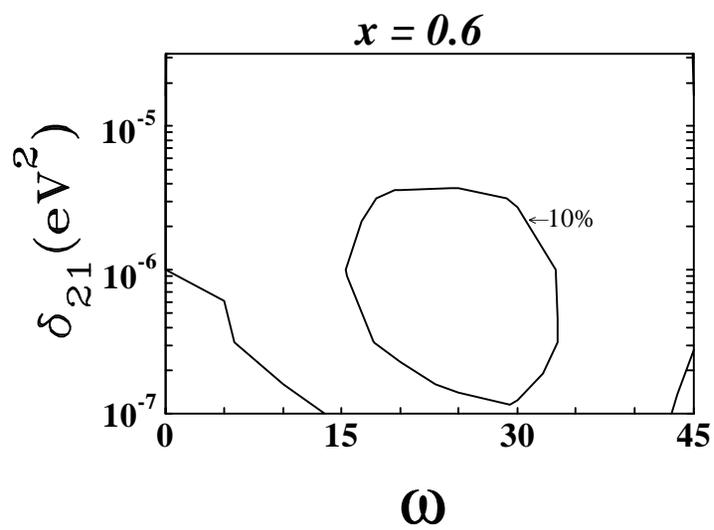
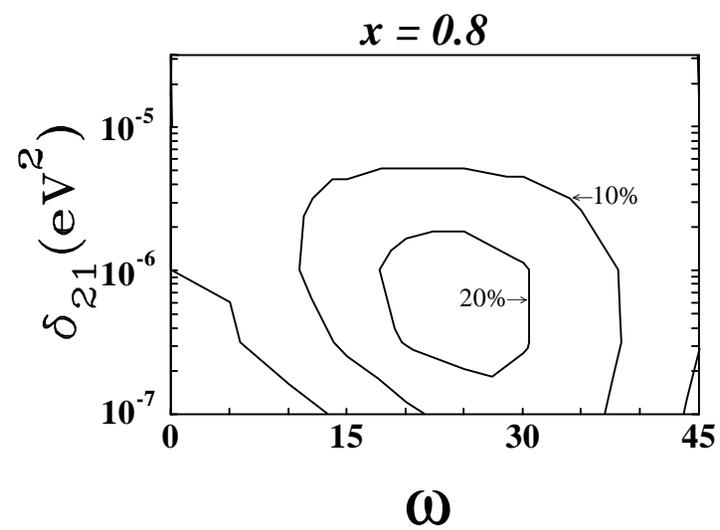
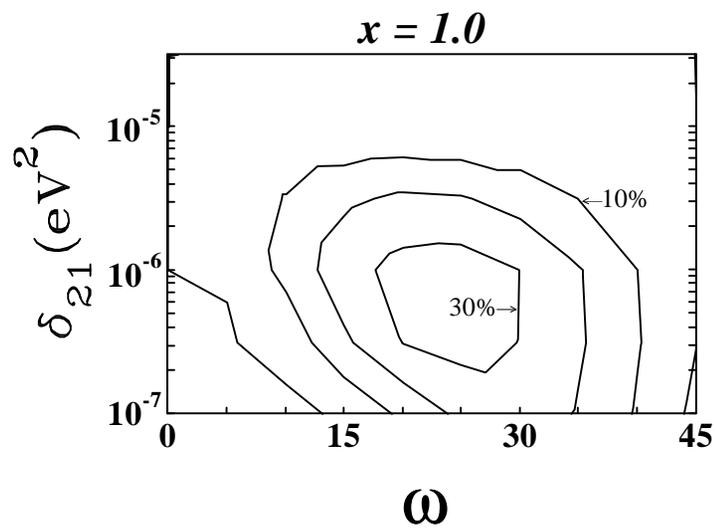
\begin{figure}
\caption{
 Contour plots of the enhancement factor for single eclipse
$F( = \displaystyle\frac {R_{M} - R_{O}} {R_{O}})$ in the $\omega - \delta_{21}$ plane
for $\phi = 30^o$ and  $x$ = 0.6, 0.8 and 1.0.
The enhancement factor (regarded as a percentage) increases by $10\%$ for
every adjacent ring , as we move inwards towards the centre of the plot.
}\label {Fig. 4}
\end{figure}
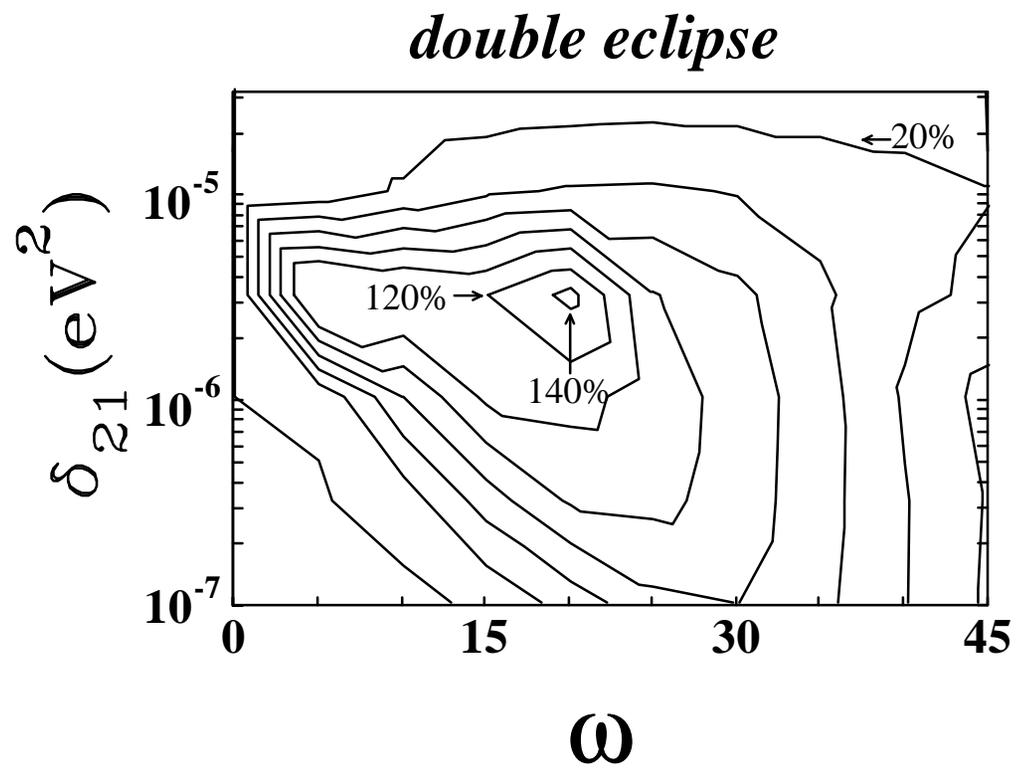
\begin{figure}
\caption{
Contour plot of the enhancement factor for double eclipse  $G (=  \displaystyle\frac {R_{M E}- R_{O}} {R_{O}}$)
for $\phi$ = 0 and $x$= 1.0. The distance travelled by the neutrino inside the
earth is also taken to be the full earth diameter. The enhancement factor
increases by $20\%$ as we move inwards. 
}\label {Fig. 5}
\end{figure}

\end{document}